\documentclass[amsmath,amssymb,showpacs,aps,prd,superscriptaddress,twocolumn,nofootinbib]{revtex4}
\usepackage{amssymb}   
\usepackage{graphicx}
\usepackage{dcolumn}
\usepackage{bm}
\usepackage{verbatim}
\usepackage{times}
\begin{document}
\title{Search for Point Sources of High Energy Neutrinos with Final Data from AMANDA-II}
\affiliation{III Physikalisches Institut, RWTH Aachen University, D-52056 Aachen, Germany}
\affiliation{Dept. of Physics and Astronomy, University of Alabama, Tuscaloosa, AL 35487, USA}
\affiliation{Dept.~of Physics and Astronomy, University of Alaska Anchorage, 3211 Providence Dr., Anchorage, AK 99508, USA}
\affiliation{CTSPS, Clark-Atlanta University, Atlanta, GA 30314, USA}
\affiliation{School of Physics and Center for Relativistic Astrophysics, Georgia Institute of Technology, Atlanta, GA 30332, USA}
\affiliation{Dept.~of Physics, Southern University, Baton Rouge, LA 70813, USA}
\affiliation{Dept.~of Physics, University of California, Berkeley, CA 94720, USA}
\affiliation{Lawrence Berkeley National Laboratory, Berkeley, CA 94720, USA}
\affiliation{Institut f\"ur Physik, Humboldt-Universit\"at zu Berlin, D-12489 Berlin, Germany}
\affiliation{Universit\'e Libre de Bruxelles, Science Faculty CP230, B-1050 Brussels, Belgium}
\affiliation{Vrije Universiteit Brussel, Dienst ELEM, B-1050 Brussels, Belgium}
\affiliation{Dept.~of Physics, Chiba University, Chiba 263-8522 Japan}
\affiliation{Dept.~of Physics and Astronomy, University of Canterbury, Private Bag 4800, Christchurch, New Zealand}
\affiliation{Dept.~of Physics, University of Maryland, College Park, MD 20742, USA}
\affiliation{Dept. of Physics and Center for Cosmology and Astro-Particle Physics, The Ohio State University, 191 W. Woodruff Ave., Columbus, OH 43210, USA}
\affiliation{Dept.~of Physics, Universit\"at Dortmund, D-44221 Dortmund, Germany}
\affiliation{Dept.~of Subatomic and Radiation Physics, University of Gent, B-9000 Gent, Belgium}
\affiliation{Max-Planck-Institut f\"ur Kernphysik, D-69177 Heidelberg, Germany}
\affiliation{Dept.~of Physics and Astronomy, University of California, Irvine, CA 92697, USA}
\affiliation{Laboratory for High Energy Physics, \'Ecole Polytechnique F\'ed\'erale, CH-1015 Lausanne, Switzerland}
\affiliation{Dept.~of Physics and Astronomy, University of Kansas, Lawrence, KS 66045, USA}
\affiliation{Dept.~of Astronomy, University of Wisconsin, Madison, WI 53706, USA}
\affiliation{Dept.~of Physics, University of Wisconsin, Madison, WI 53706, USA}
\affiliation{Institute of Physics, University of Mainz, Staudinger Weg 7, D-55099 Mainz, Germany}
\affiliation{University of Mons-Hainaut, 7000 Mons, Belgium}
\affiliation{Bartol Research Institute and Department of Physics and Astronomy, University of Delaware, Newark, DE 19716, USA}
\affiliation{Dept.~of Physics, University of Oxford, 1 Keble Road, Oxford OX1 3NP, UK}
\affiliation{Dept.~of Physics, University of Wisconsin, River Falls, WI 54022, USA}
\affiliation{Dept.~of Physics, Stockholm University, SE-10691 Stockholm, Sweden}
\affiliation{Dept.~of Astronomy and Astrophysics, Pennsylvania State University, University Park, PA 16802, USA}
\affiliation{Dept.~of Physics, Pennsylvania State University, University Park, PA 16802, USA}
\affiliation{Division of High Energy Physics, Uppsala University, S-75121 Uppsala, Sweden}
\affiliation{Dept.~of Physics and Astronomy, Utrecht University/SRON, NL-3584 CC Utrecht, The Netherlands}
\affiliation{Dept.~of Physics, University of Wuppertal, D-42119 Wuppertal, Germany}
\affiliation{DESY, D-15735 Zeuthen, Germany}

\author{R.~Abbasi}
\affiliation{Dept.~of Physics, University of Wisconsin, Madison, WI 53706, USA}
\author{M.~Ackermann}
\affiliation{DESY, D-15735 Zeuthen, Germany}
\author{J.~Adams}
\affiliation{Dept.~of Physics and Astronomy, University of Canterbury, Private Bag 4800, Christchurch, New Zealand}
\author{M.~Ahlers}
\affiliation{Dept.~of Physics, University of Oxford, 1 Keble Road, Oxford OX1 3NP, UK}
\author{J.~Ahrens}
\affiliation{Institute of Physics, University of Mainz, Staudinger Weg 7, D-55099 Mainz, Germany}
\author{K.~Andeen}
\affiliation{Dept.~of Physics, University of Wisconsin, Madison, WI 53706, USA}
\author{J.~Auffenberg}
\affiliation{Dept.~of Physics, University of Wuppertal, D-42119 Wuppertal, Germany}
\author{X.~Bai}
\affiliation{Bartol Research Institute and Department of Physics and Astronomy, University of Delaware, Newark, DE 19716, USA}
\author{M.~Baker}
\affiliation{Dept.~of Physics, University of Wisconsin, Madison, WI 53706, USA}
\author{B.~Baret}
\affiliation{Vrije Universiteit Brussel, Dienst ELEM, B-1050 Brussels, Belgium}
% ...

\date{\today}

\collaboration{IceCube Collaboration}
\noaffiliation


\begin{abstract}
We present a search for point sources of high energy neutrinos using 3.8 years of data recorded by AMANDA-II during 2000-2006.
After reconstructing muon tracks and applying selection criteria designed to optimally retain neutrino-induced events originating
in the Northern Sky, we arrive at a sample of 6595 candidate events, predominantly from atmospheric neutrinos with primary energy 100 GeV to 8 TeV.
Our search of this sample reveals no indications of a neutrino point source. We place the most stringent limits to date on $E^{-2}$ neutrino fluxes from
points in the Northern Sky, with an average upper limit of $E^{2}\Phi_{\nu_{\mu} + \nu_{\tau}} \le$ 5.2~$\times$~10$^{-11}$ TeV
cm$^{-2}$ s$^{-1}$ on the sum of $\nu_{\mu}$ and $\nu_{\tau}$ fluxes, assumed equal, over the energy range from 1.9 TeV to 2.5 PeV.
\end{abstract}
\maketitle
\section{Introduction}
Detecting extraterrestrial sources of high energy ($>$TeV) neutrinos is a longstanding goal of astrophysics.  Neutrinos
are neither deflected by magnetic fields nor significantly attenuated by matter and radiation en route to Earth,
thus neutrino astronomy offers an undistorted view deep into the high energy universe.  Particularly, neutrinos offer an opportunity
to probe the sources of high energy cosmic rays, which remain unknown.  Potential cosmic ray sources include
galactic microquasars and supernova remnants as well as extragalactic sources such
as active galactic nuclei and gamma ray bursts.  These objects are thought to accelerate protons and nuclei
in shock fronts via the Fermi mechanism \cite{fermi}, resulting in power law energy spectra $E^{\alpha}$, with $\alpha$ $\sim$ $-$2.
A fraction of the energized particles interact with local matter and radiation, producing pions.  The neutral
pions decay into high energy photons, and the charged pions ultimately produce neutrinos with a flavor ratio
$\nu_e$:$\nu_{\mu}$:$\nu_{\tau}$ $\sim$1:2:0, mixing to approximately 1:1:1 at Earth because of vacuum flavor oscillations.
Observations of TeV gamma rays \cite{milagrogal,hessgal,magicgal} hint at possible cosmic ray source locations but
currently cannot separate neutral pion decay spectra from inverse Compton emission.  The Auger collaboration has reported
a correlation of arrival directions of the highest energy cosmic rays with active galactic
nuclei \cite{augercorrsci}; however, a similar correlation has not been observed by HiRes \cite{hiresehe}.
Identification of a high energy neutrino point source would provide an unambiguous signature of energetic hadrons and cosmic
ray acceleration.  Neutrino flux predictions exist for many potential sources \cite{microq5039,microqlsi,snr,stecker,kappes,beacom},
but no high energy neutrino point source has yet been identified \cite{amandalim, macrolim, superklim}.

The search for high energy neutrino point sources is a major objective of the Antarctic Muon And Neutrino Detector Array
(AMANDA).  High energy leptons are produced in the Earth by charged-current neutrino interactions.  In transparent
matter, a cone of Cherenkov photons propagates from the lepton track according to the optical properties of the medium.
AMANDA-II is an optical Cherenkov detector consisting of 677 optical modules arranged in 19 strings frozen $\sim$1500 m to
$\sim$2000 m deep in the ice sheet at the geographic South Pole.
Approximately 540 modules in the core of the array showing stable performance are used in this search.
Each module contains a 20 cm diameter photomultiplier tube (PMT) optically coupled to an outer glass high-pressure sphere.  PMT pulses are propagated
to surface electronics, and, when the trigger threshold of 24 discriminator crossings (``hits") within 2.5 $\mu$s is satisfied, the pulse
leading edge times are recorded.  The leading edge times along with known detector geometry and optical properties of South Pole ice \cite{icepaper}
allow reconstruction of tracks passing through the detector \cite{amandareco}.  High energy electrons produce short electromagnetic
cascades with little directional information of the primary neutrino.  Muons produced in
the ice and bedrock, on the other hand, propagate up to several kilometers to the detector and their tracks are reconstructed with ~1.5$^{\circ}$--2.5$^{\circ}$
median accuracy depending on energy and zenith angle.
Tau leptons decay rapidly and produce tracks too short for reconstruction
below $\sim$PeV energies.  Tau decay, however, contributes high energy muons with a branching ratio of 17.7\% \cite{amandalim, pdg},
and these muon tracks can be reconstructed.  We thus search for upward propagating muons produced in the Earth by $\nu_{\mu}$ ($\bar{\nu}_{\mu}$) and
$\nu_{\tau}$ ($\bar{\nu}_{\tau}$) fluxes
following roughly an $E^{-2}$ energy spectrum.  While downward neutrino induced muons also trigger the detector, such events
are difficult to distinguish from downward muons produced by cosmic ray air showers.  Located at the
South Pole, AMANDA-II is thus most sensitive to neutrino fluxes from the Northern Sky.  Air showers also produce neutrinos,
and this atmospheric neutrino flux \cite{bartol,honda} is the main background for our search.

Here we present the results of a search for astrophysical point sources of high energy neutrinos
using 3.8 years of data recorded by AMANDA-II during 2000-2006, extending the previous five-year analysis \cite{amandalim}
with data from the final two years of standalone operation and improving our sensitivity by a factor of $\sim$2.
We report flux limits for a catalog of 26 selected source candidates along with results of a search for neutrino
sources over the entire Northern Sky.  Additionally, we report results from a search for
neutrino emission from gamma ray sources identified by Milagro \cite{milagrogal} and a search for event angular correlations.
In all cases, we observe no indications of an astrophysical neutrino point source.

\section{Data Selection}

\begin{figure}\begin{center}
\mbox{\includegraphics[width=8.6cm]{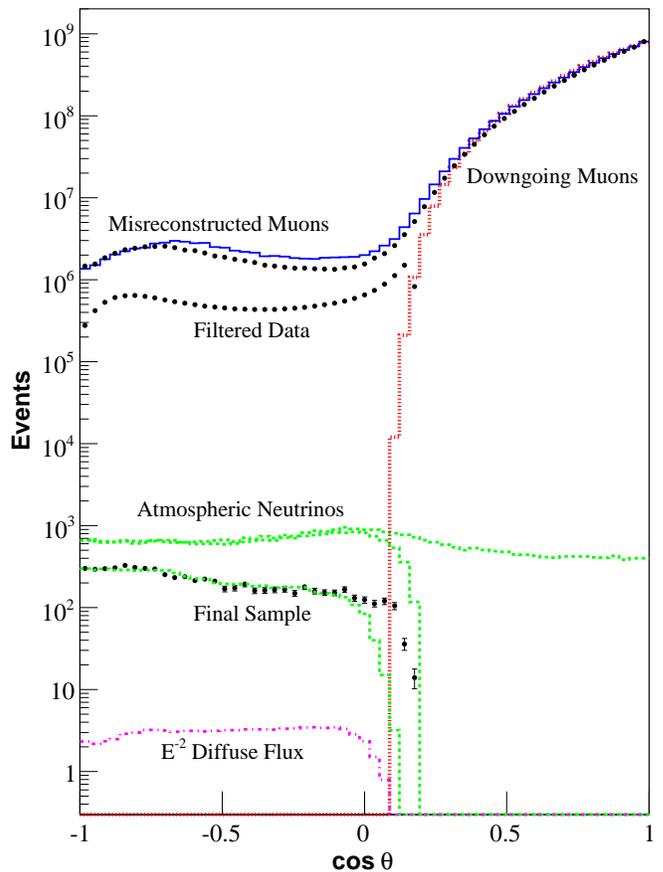}}
\caption{\label{Fig:Zall} Zenith angle ($\theta$) distributions for data and simulation at several reduction levels.  Reconstructed (solid) and true
(fine dotted) zenith angle distributions are shown for CORSIKA \cite{corsika} cosmic ray muon simulation at retrigger level, and reconstructed zenith angle
distributions are shown for atmospheric neutrino simulation (dotted) and data (circles) at retrigger level, filter level, and final selection.
We also show the reconstructed zenith angle distribution of a diffuse $E^{-2}$ neutrino flux at the current limit \cite{diffuse} using our final selection
(dash-dotted).}
\end{center}\end{figure}

As illustrated in Fig.~\ref{Fig:Zall}, AMANDA-II records $O(10^9)$ events per year from downward propagating muons produced by cosmic ray air showers, $O(10^3)$
events per year from atmospheric neutrinos, and $O(10)$ high quality events per year from astrophysical $E^{-2}$ neutrino fluxes
given current limits \cite{diffuse}.  We attempt to isolate these neutrino events from the
downward muon background in a computationally
efficient manner.  We exclude data taken during periods of detector instability and significant maintenance, which include the austral summer
(November 1 through February 15).  After accounting for deadtime in data acquisition electronics, nominally $\sim$15\% of uptime, we have accumulated
1387 days (3.8 years) of livetime with 1.29$\times 10^{10}$ events during seven years of operation (table \ref{Tab:Esel}).
\begin{table}[t!!!]
\begin{tabular}{lcccccc}
\hline\hline
Year & Livetime & Total Events && Filtered Events && Final Selection\\
\hline
2000 & 197 d & 1.37$\times 10^9$ && 1.63$\times 10^6$ && 596\\
2001 & 193 d & 2.00$\times 10^9$ && 1.90$\times 10^6$ && 854\\
2002 & 204 d & 1.91$\times 10^9$ && 2.10$\times 10^6$ && 1009\\
2003 & 213 d & 1.86$\times 10^9$ && 2.22$\times 10^6$ && 1069\\
2004 & 194 d & 1.72$\times 10^9$ && 2.09$\times 10^6$ && 998\\
2005 & 199 d & 2.06$\times 10^9$ && 5.21$\times 10^6$ && 1019\\
2006 & 187 d & 2.00$\times 10^9$ && 4.89$\times 10^6$ && 1050\\ & & && && \\
\bf{Total} & 1387 d & 12.92$\times 10^9$ && 20.04$\times 10^6$ && 6595\\
\hline\hline
\end{tabular}
\caption{\label{Tab:Esel}
AMANDA livetime and event totals.}
\end{table}

Events are first processed to remove hits induced by electrical cross talk, hits from unstable modules, and isolated noise hits \cite{amandareco}, and events
which no longer pass the trigger criteria are discarded.  These retriggered events
are then reconstructed with the fast pattern matching algorithms DirectWalk (DW) \cite{amandareco}
and JAMS \cite{jams} which identify muon tracks within events.  For optimal efficiency, our upgoing event selection requires both zenith angles $\theta_{DW}$
and $\theta_{JAMS}$ greater than $70^{\circ}$--$80^{\circ}$.
\begin{figure*}[th]\begin{center}
\begin{tabular}{l@{\hspace{1cm}}r}
\mbox{\includegraphics[width=7.3cm]{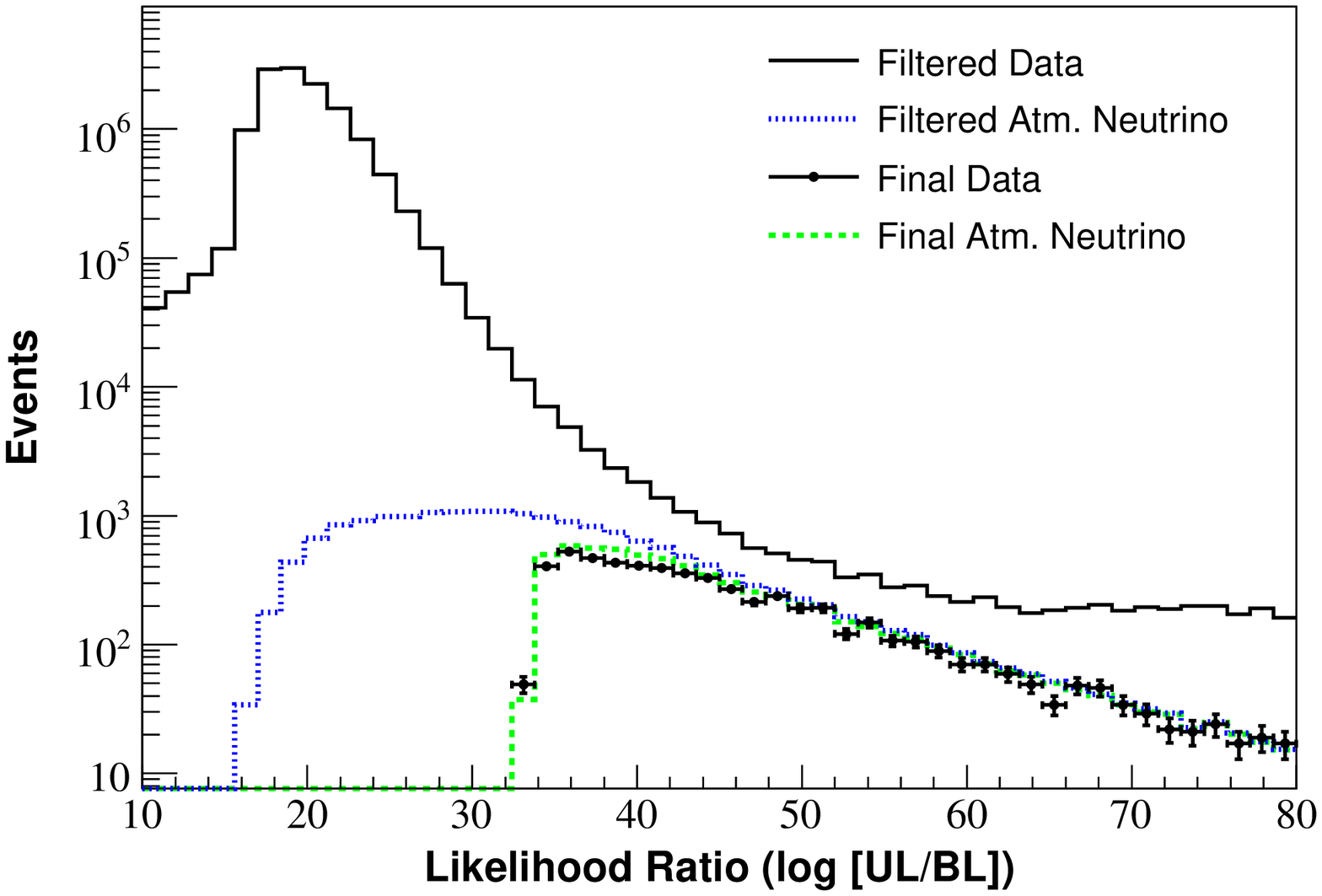}} & \mbox{\includegraphics[width=7.3cm]{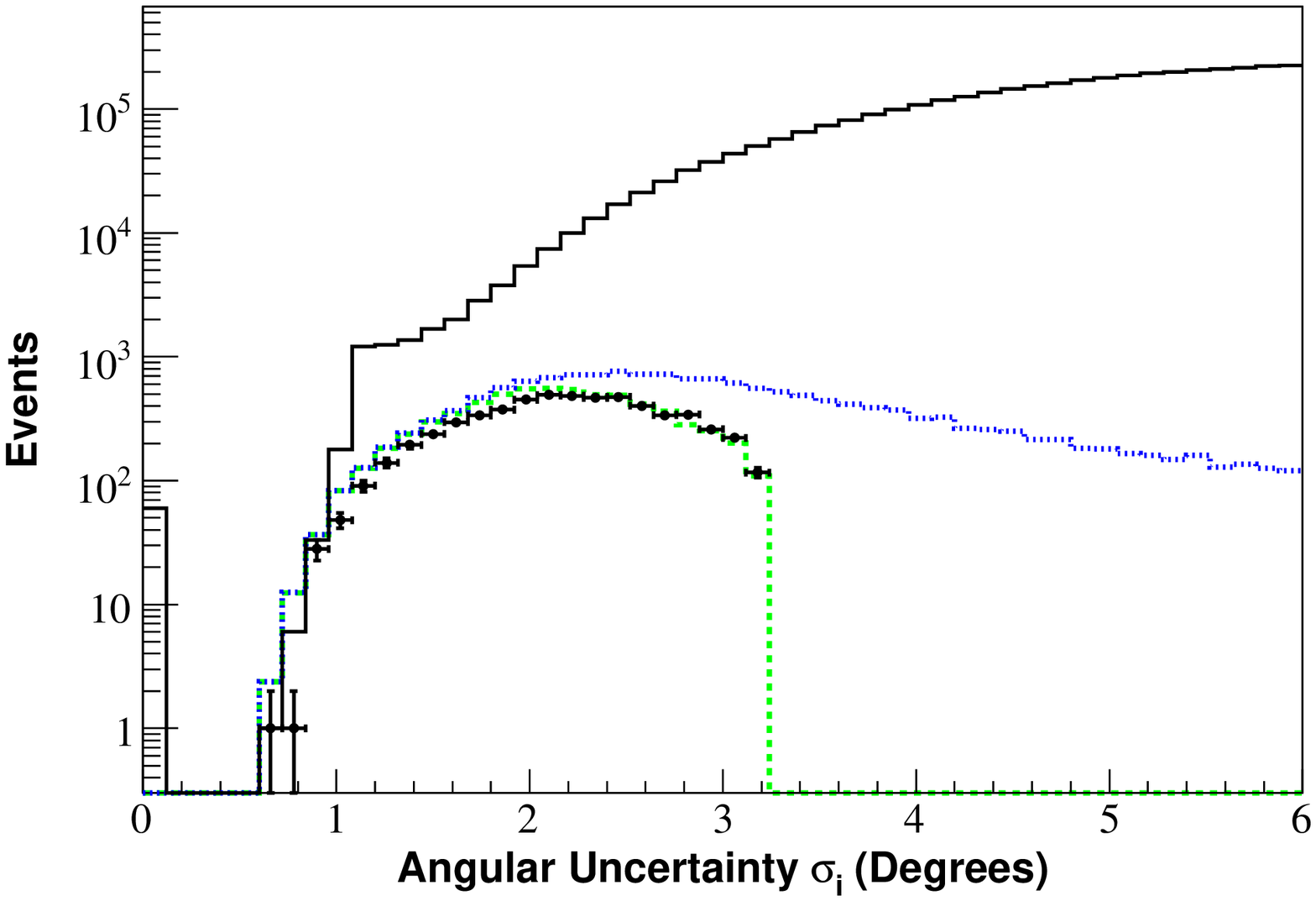}} \\
\mbox{\includegraphics[width=7.3cm]{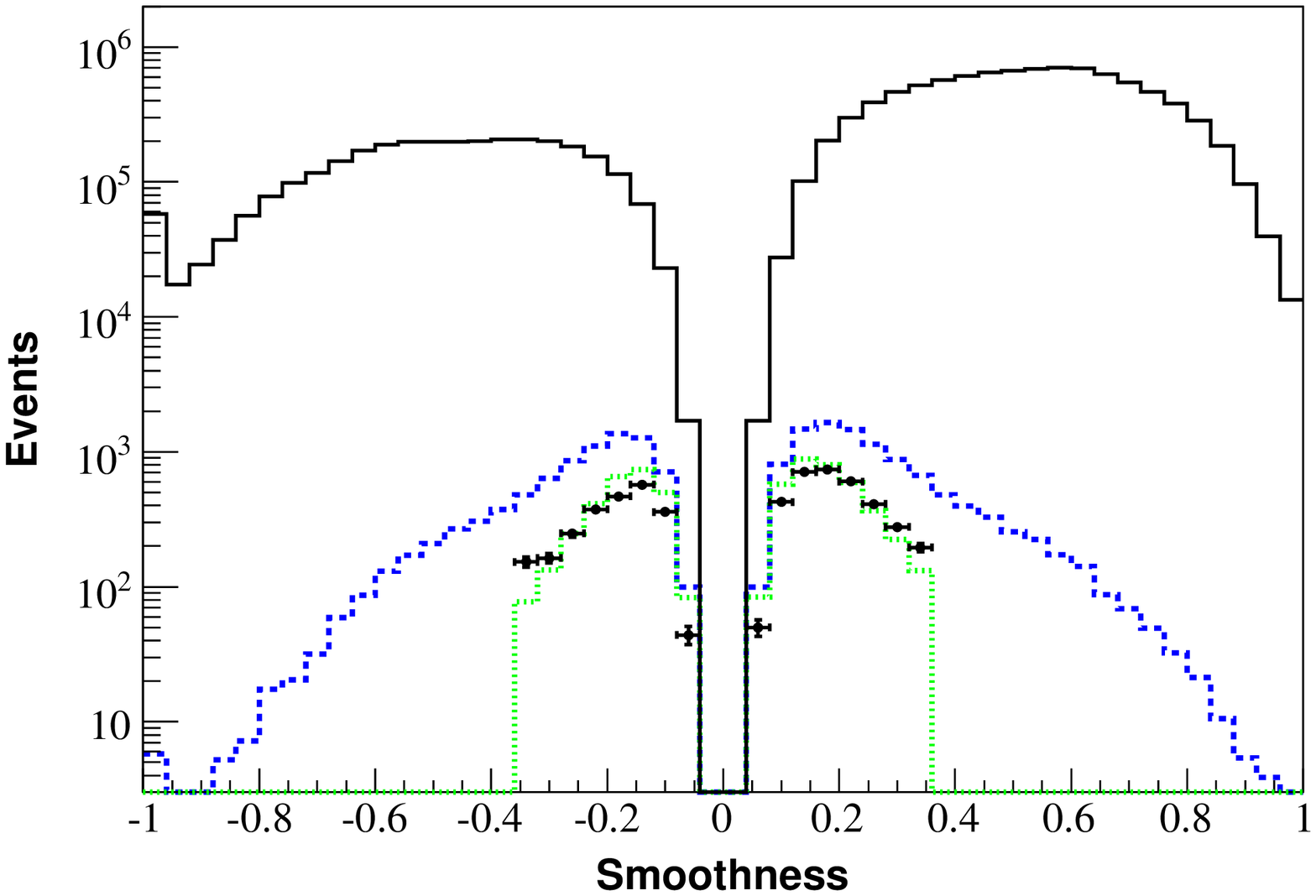}} & \mbox{\includegraphics[width=7.3cm]{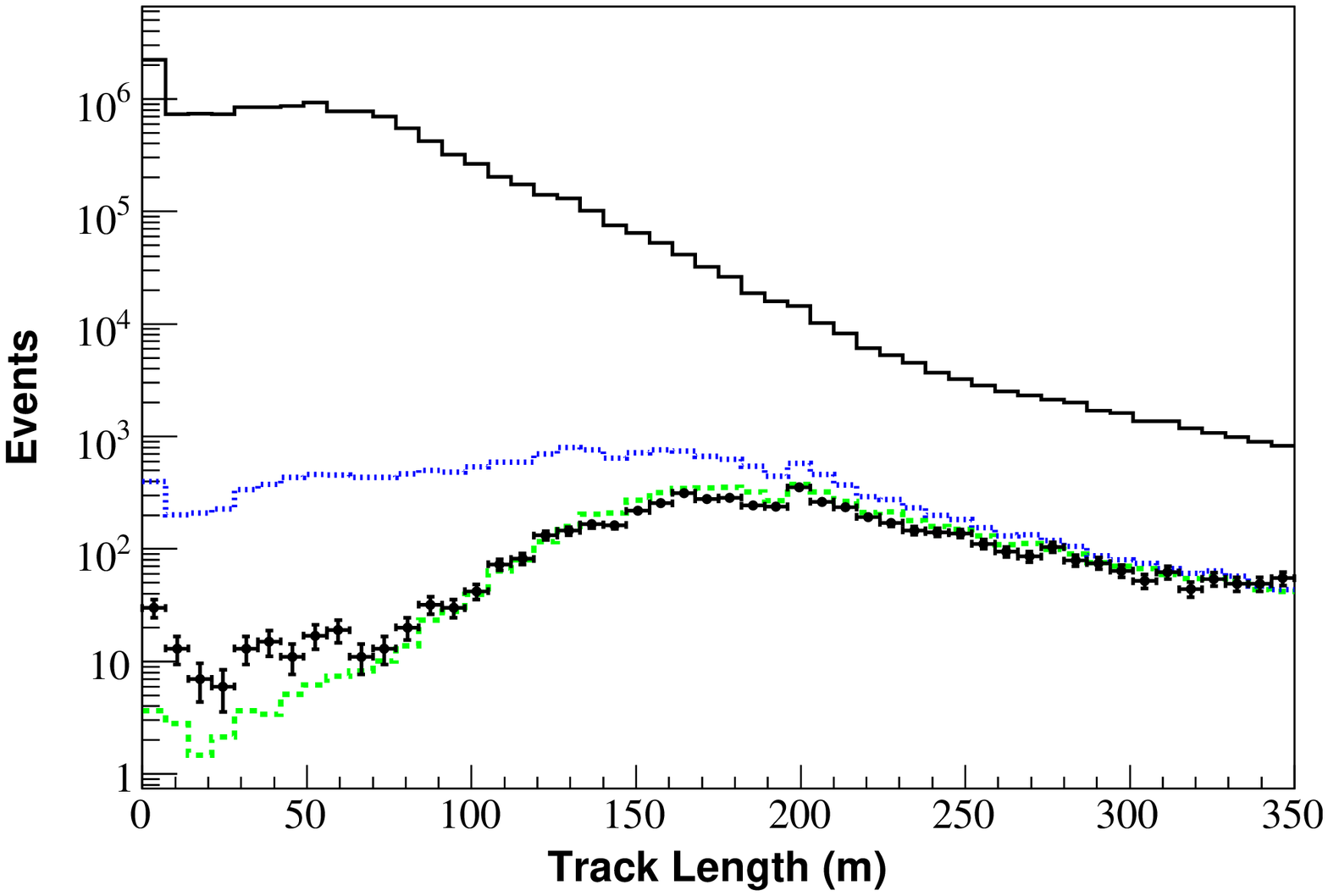}} \\
\mbox{\includegraphics[width=7.3cm]{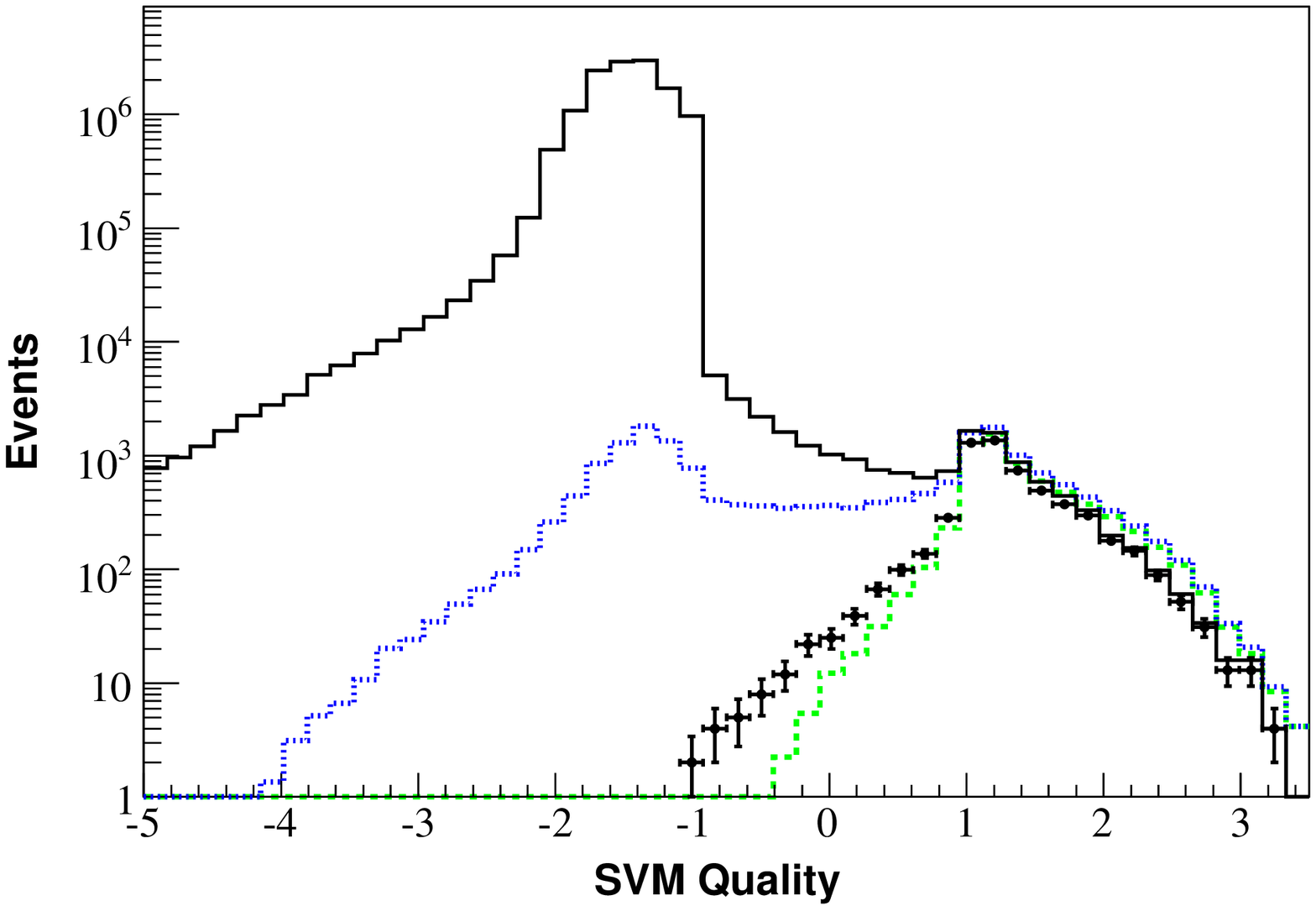}} & \mbox{\includegraphics[width=7.3cm]{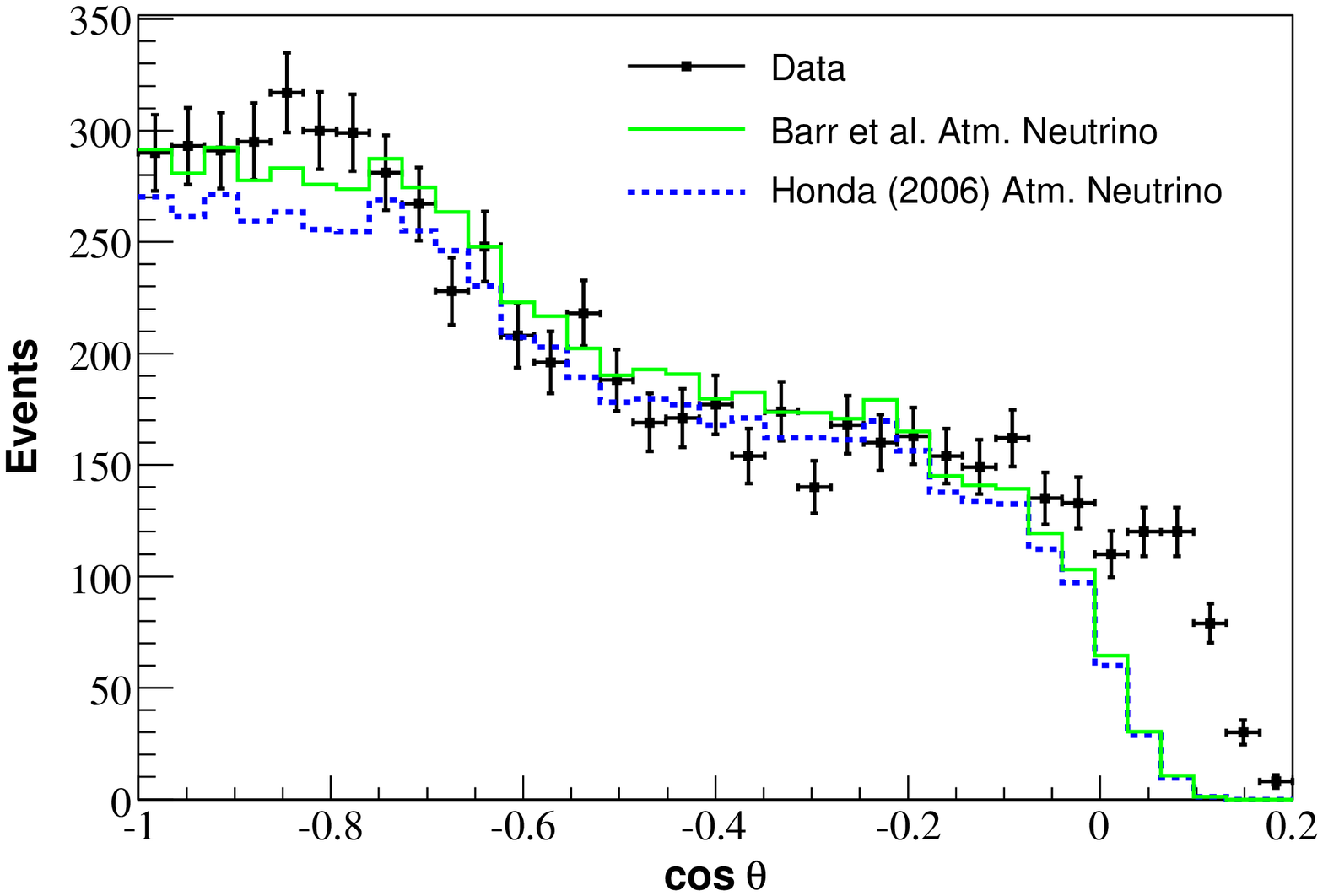}} \\
\end{tabular}
\caption{\label{Fig:Par} Distributions of data and atmospheric neutrinos at filter level and final selection level for several parameters and zenith
angles $\theta > 95^{\circ}$ (top and left),
and zenith angle distribution for the selected 6595 neutrino candidate events compared
with model predictions \cite{bartol,honda} for atmospheric neutrinos (bottom right).}
\end{center}\end{figure*}
\begin{figure*}[tb]\begin{center}
\mbox{\includegraphics[width=14cm]{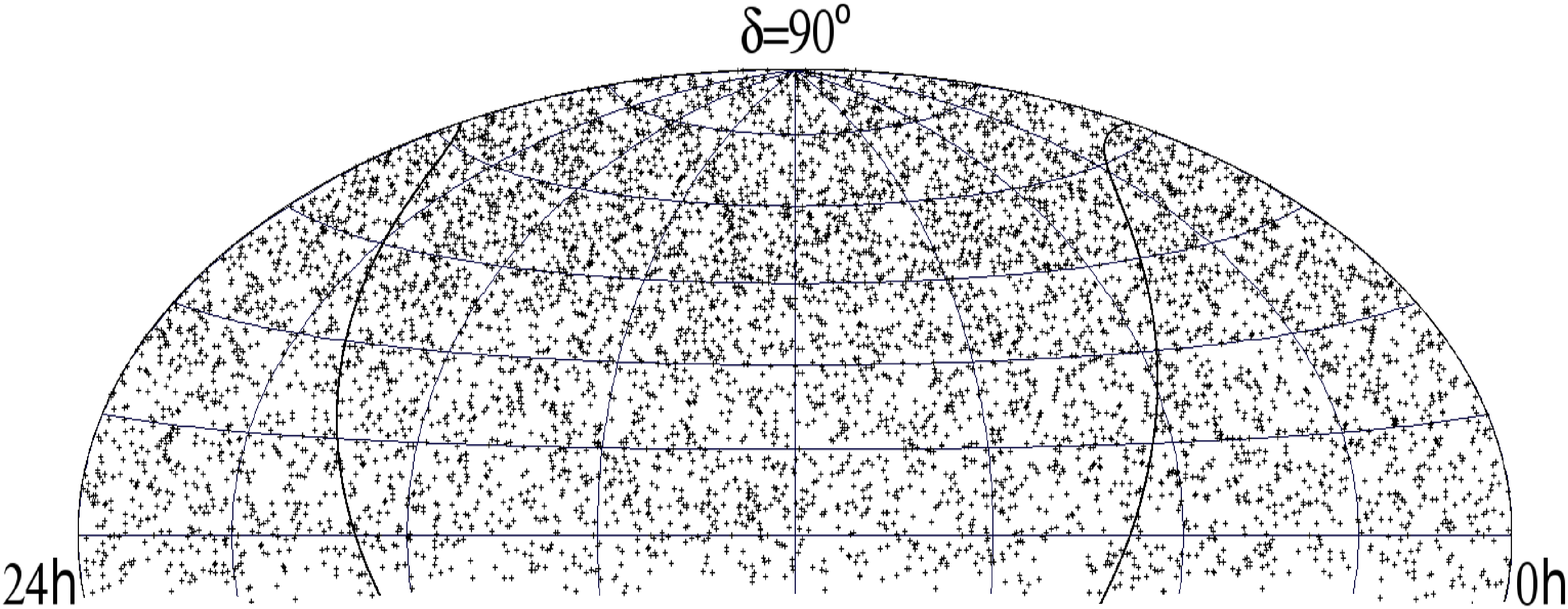}}
\caption{\label{Fig:Emap} Equatorial sky map of 6595 events recorded by
AMANDA-II from 2000--2006.  A table of the events is available \cite{events}.}
\end{center}\end{figure*}

Two CPU intensive maximum likelihood reconstructions are applied to events passing the above selection criteria.  First, we apply an
unbiased likelihood (UL) fit seeded with the DirectWalk and JAMS reconstructed tracks and 30 additional random track directions.  The likelihood
function \cite{amandareco} parameterizes the probability of observing the obtained geometry and leading edge times of hit modules in terms
of track zenith angle, azimuthal angle, and position.  The likelihood is maximized with respect to these parameters (in practice, the negative logarithm of the
likelihood is numerically minimized), yielding the best fit track zenith and azimuthal angles,
and the fit result from the seed yielding the maximum likelihood is chosen as the reconstructed track.  A 64 seed Bayesian likelihood (BL) fit is also
done, using the downgoing muon zenith angle distribution as a Bayesian prior.  With the additional
cut $\theta_{UL} > 80^{\circ}$, our upgoing event filter reduces the downward muon background by
a factor of $\sim$650 relative to trigger level (table \ref{Tab:Esel}).

After this cut, $O(10^6)$ misreconstructed downward muon events per year remain, which still outnumber atmospheric neutrinos
by roughly three orders of magnitude.  The vast majority of these events are removed by the following four topological parameters, shown in Fig. \ref{Fig:Par}:
\begin{itemize}
\item The likelihood ratio of the UL and BL fits.  Downgoing muon background events misreconstructed as upgoing by the UL fit typically are also fit well
with the downward biased BL reconstruction, whereas true upgoing events are not.  Therefore, the UL/BL likelihood ratio tends to be higher for upgoing events.
\item The angular uncertainty of the UL fit, described further in section III.  Misreconstructed events generally have large angular uncertainty.
\item The {\it smoothness}, or homogeneity of the hit distribution along the UL track \cite{amandareco}.  High quality events contain photon hits along the
entire length of the track and have smoothness values near zero, whereas hits from misreconstructed events tend to distribute near the beginning or
end of the track and have smoothness values near +1 and $-$1, respectively.
\item The UL track {\it direct length}, obtained by projecting {\it direct hits} backward to the UL track at the
Cherenkov angle and taking the distance along the track between the first and last.  We select direct hits, compatible with relatively unscattered photons
and arriving on-time with the Cherenkov cone, using the time window $-15$~ns~$<~t-t_{ch}~<~25$~ns \cite{amandareco}.  Hits from misreconstructed
events rarely follow the muon-Cherenkov timing pattern over significant distances, resulting in short lengths.
\end{itemize}
For the zenith angle region $91.5^{\circ} < \theta < 180^{\circ}$ we use the following zenith angle dependent cuts, optimized
to yield maximum sensitivity \cite{mrf}:
\begin{center}\begin{tabular}{rcl}
$\log(UL/BL)$&$>$&$34 - 25\cdot \Phi(\cos~\theta + 0.15)$\\
$\sigma_i$&$<$&$3.2 - 4\cdot \Phi(-\cos~\theta - 0.75)$\\
$|$Smoothness$|$&$<$&$0.36$.\\
\end{tabular}\end{center}
Here $\Phi(x) = x$ for positive $x$, and $\Phi(x) = 0$ for $x < 0$.
We use a support vector machine (SVM) \cite{svmlight} trained on the four parameters to improve event selection in the near-horizontal region
80$^{\circ} < \theta < 91.5^{\circ}$. Events with SVM quality of zero or less are consistent with misreconstructed muon background, while events with
larger values of SVM quality are increasingly consistent with quality muons. We apply the cut:
\begin{center}\begin{tabular}{rcl}
SVM Quality &$>$&$1 - 12\cdot \Phi(\cos~\theta - 0.023)$\\
\end{tabular}\end{center}
Application of these quality cuts yields 6595 neutrino candidate events \cite{events} (Fig. \ref{Fig:Emap}).

\begin{figure}\begin{center}
\mbox{\includegraphics[width=8.6cm]{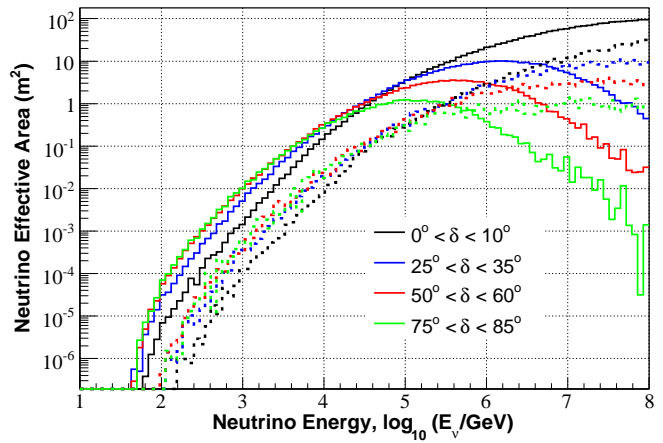}}
\caption{\label{Fig:Aeff} Effective area for averaged $\nu_{\mu}$ and $\bar{\nu}_{\mu}$ (solid) and averaged $\nu_{\tau}$ and $\bar{\nu}_{\tau}$
(dashed) neutrino fluxes for several declination ranges.}
\end{center}\end{figure}

\begin{figure}\begin{center}
\mbox{\includegraphics[width=8.6cm]{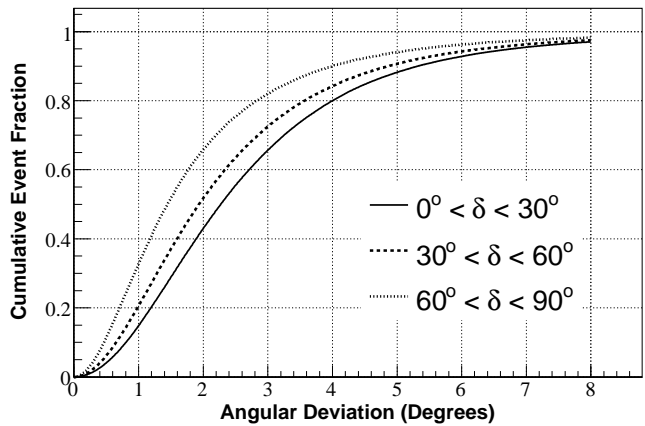}}
\caption{\label{Fig:Psi} Angular deviation between neutrino and UL fit track for simulated $E^{-2}$ muon neutrino events
from several declination ranges.}
\end{center}\end{figure}

Simulations of two atmospheric neutrino flux models \cite{bartol,honda}, with events generated by ANIS \cite{anis} and resultant muons propagated
to the detector with MMC \cite{mmc}, both agree with data in track quality parameter distributions and zenith angle (Fig. \ref{Fig:Par}) within
the $\sim$30\% uncertainty in these flux predictions.  Application of the filter
selection and final quality cuts to this simulation yields an atmospheric neutrino efficiency of 30\% relative to retrigger level
for $\theta > 90^{\circ}$.
The contribution of misreconstructed downward muons has been estimated by subtracting the simulated atmospheric neutrino rate, after
renormalizing it for a more stringent selection yielding a nearly pure neutrino sample.  The muon contamination has been found to be
less than 5\% for $\theta> 95^{\circ}$ (declination $\delta >$ 5$^{\circ}$), but the contamination is more significant near the equator
and dominates events in the Southern Sky.  A parallel analysis of these
atmospheric neutrino events has revealed no evidence of new physics such as violation of Lorentz invariance and quantum decoherence \cite{jkelley}.
We simulate $\nu_{\mu}$ and $\nu_{\tau}$ events from 10 GeV to 100 PeV with an identical software chain, and this simulation is used to calculate
neutrino effective area, shown in Fig. \ref{Fig:Aeff}, and flux limits for neutrino sources with $E^{-2}$ energy spectra.  The central 90\% of such
signal events fall within the energy range 1.9 TeV to 2.5 PeV.  The median accuracy of the
UL fit when applied to simulated events following an $E^{-2}$ energy spectrum is 1.5$^{\circ}$--2.5$^{\circ}$, shown in Fig. \ref{Fig:Psi}.
The absolute pointing accuracy of AMANDA has been confirmed by observing downgoing muon events coincident with well-reconstructed air showers recorded by
SPASE \cite{amandareco} and events coincident with IceCube.

\section{Search Method}

\begin{figure}\begin{center}
\mbox{\includegraphics[width=8.6cm]{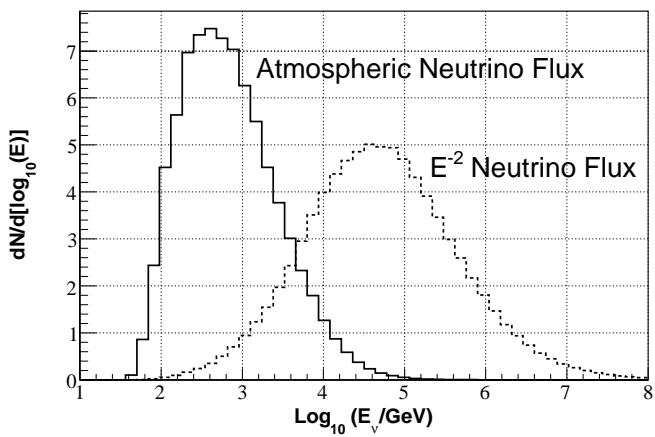}}
\caption{\label{Fig:EdistRaw} Energy distribution of events passing selection criteria for simulated atmospheric neutrino background \cite{bartol}
in a 3.5$^{\circ}$ bin and an $E^{-2}$ point source with flux $\Phi_{\nu_{\mu} + \nu_{\tau}} = 10^{-10}$ TeV cm$^{-2}$ s$^{-1}$.  Such a source
would be detected at 5$\sigma$ in approximately 40\% of trials.}
\end{center}\end{figure}

The remaining background, mostly atmospheric neutrinos, is difficult to reduce further without significantly decreasing signal efficiency.
Neutrinos from $E^{-2}$ sources are typically more energetic than atmospheric neutrinos (Fig. \ref{Fig:EdistRaw}), which follow a steeper $\sim$$E^{-3.7}$
energy spectrum.  We search our sample of 6595 events for excesses above the atmospheric neutrino background both in direction and event energy using
an unbinned maximum likelihood search method \cite{method}, providing direction and energy discrimination on an event-by-event basis by incorporating
an event angular resolution estimate and energy estimate.

\subsection{Event Angular Uncertainty Estimation}

\begin{figure}\begin{center}
\mbox{\includegraphics[width=8.6cm]{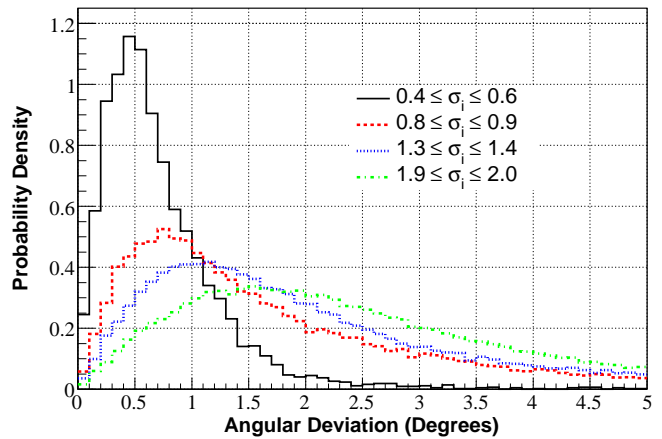}}
\caption{\label{Fig:Perr} Distributions of angular deviation between true and reconstructed tracks for simulated neutrino events
over several ranges of estimated angular uncertainty.}
\end{center}\end{figure}

Our ability to reconstruct muon tracks in AMANDA partially depends on event topology.  A muon track passing through a larger portion of the
detector or giving hits in a larger number of modules should, on average, reconstruct with better angular resolution due to a
longer lever-arm or larger number of measurement points, respectively.  We therefore estimate the resolution of each UL track by evaluating the
likelihood space near the maximum \cite{till}.  As the track zenith angle
and azimuthal angle coordinates ($\theta$, $\phi$) move away from the best fit track values ($\hat{\theta}$, $\hat{\phi}$), the quantity
$\log\mathcal{L}$ decreases parabolically from its maximum.
The likelihood ratio $-2\cdot \log\big(\frac{\mathcal{L}(\theta, \phi)}{\mathcal{L}(\hat{\theta}, \hat{\phi})}\big)$
is evaluated on a grid of zenith and azimuthal angles near the best track, and the resulting values are fit to a paraboloid with the form
\begin{equation}
-2\cdot \log\Big(\frac{\mathcal{L}(\theta, \phi)}{\mathcal{L}(\hat{\theta}, \hat{\phi})}\Big) = 
\frac{x^2}{\sigma^2_x} + \frac{y^2}{\sigma^2_y},
\end{equation}
where the $x$ and $y$ axes are fit and do not necessarily correspond to zenith and azimuthal angles.  The two errors $\sigma_x$ and $\sigma_y$ are
then geometrically averaged into a single, circular error $\sigma_i$.  The paraboloid fit is thus a convenient approximation of the likelihood
space, reducing the complex map of $-2\cdot \log\big(\frac{\mathcal{L}(\theta, \phi)}{\mathcal{L}(\hat{\theta}, \hat{\phi})}\big)$
into just $\sigma_i$.  The corresponding spatial probability density estimate at an angular distance $\Psi$ is then:
\begin{equation}
P(\Psi_i) =
\frac{e^{-\frac{\Psi^2}{2\sigma^2_i}}}{2\pi \sigma_i^2}.
\end{equation}
Distributions of the angular deviation between true and reconstructed neutrino tracks for several ranges of estimated angular
uncertainty (Fig. \ref{Fig:Perr}) show the correlation between estimated angular uncertainty and track reconstruction error.

\subsection{Event Energy Estimation}

\begin{figure}\begin{center}
\mbox{\includegraphics[width=8.6cm]{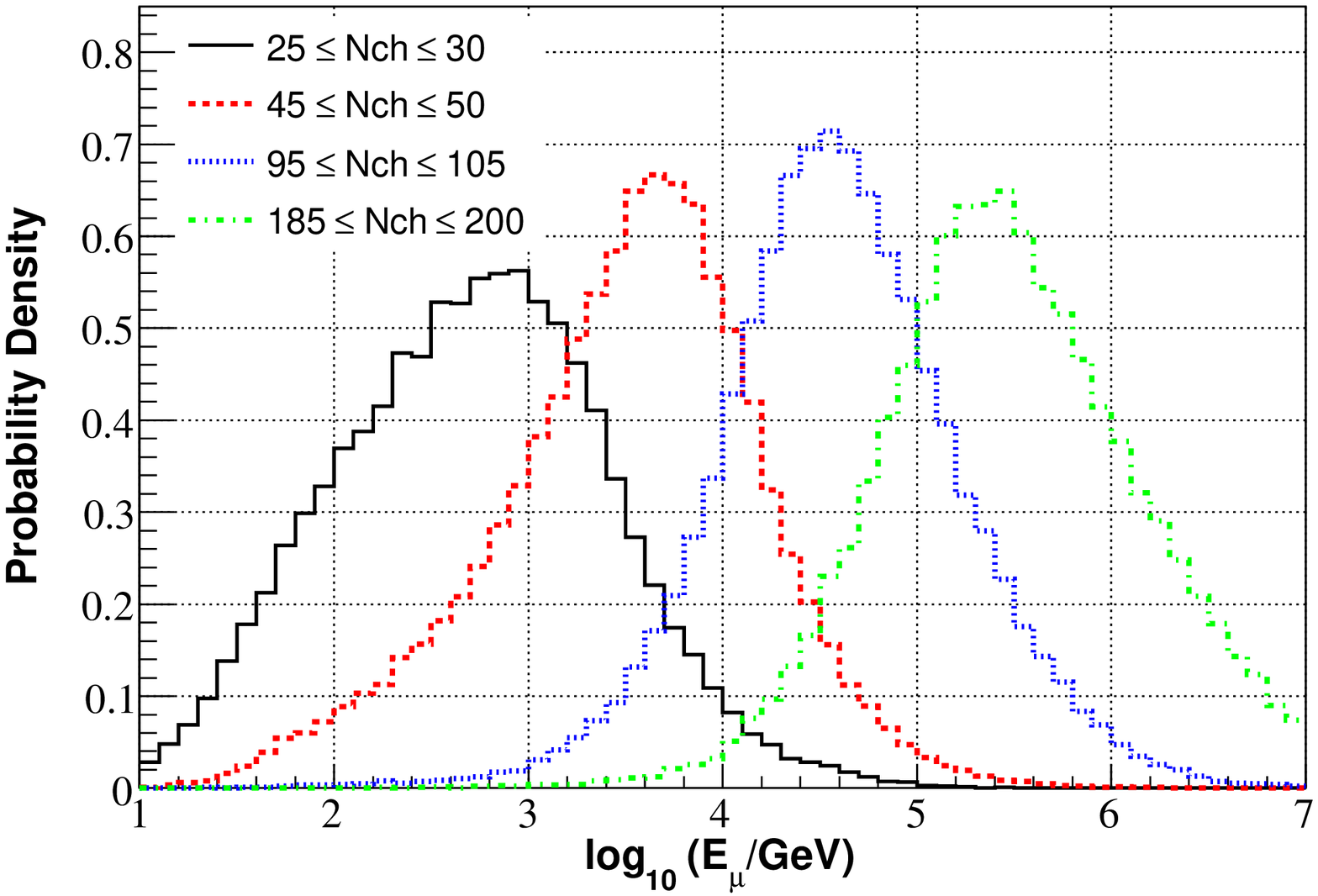}}
\mbox{\includegraphics[width=8.6cm]{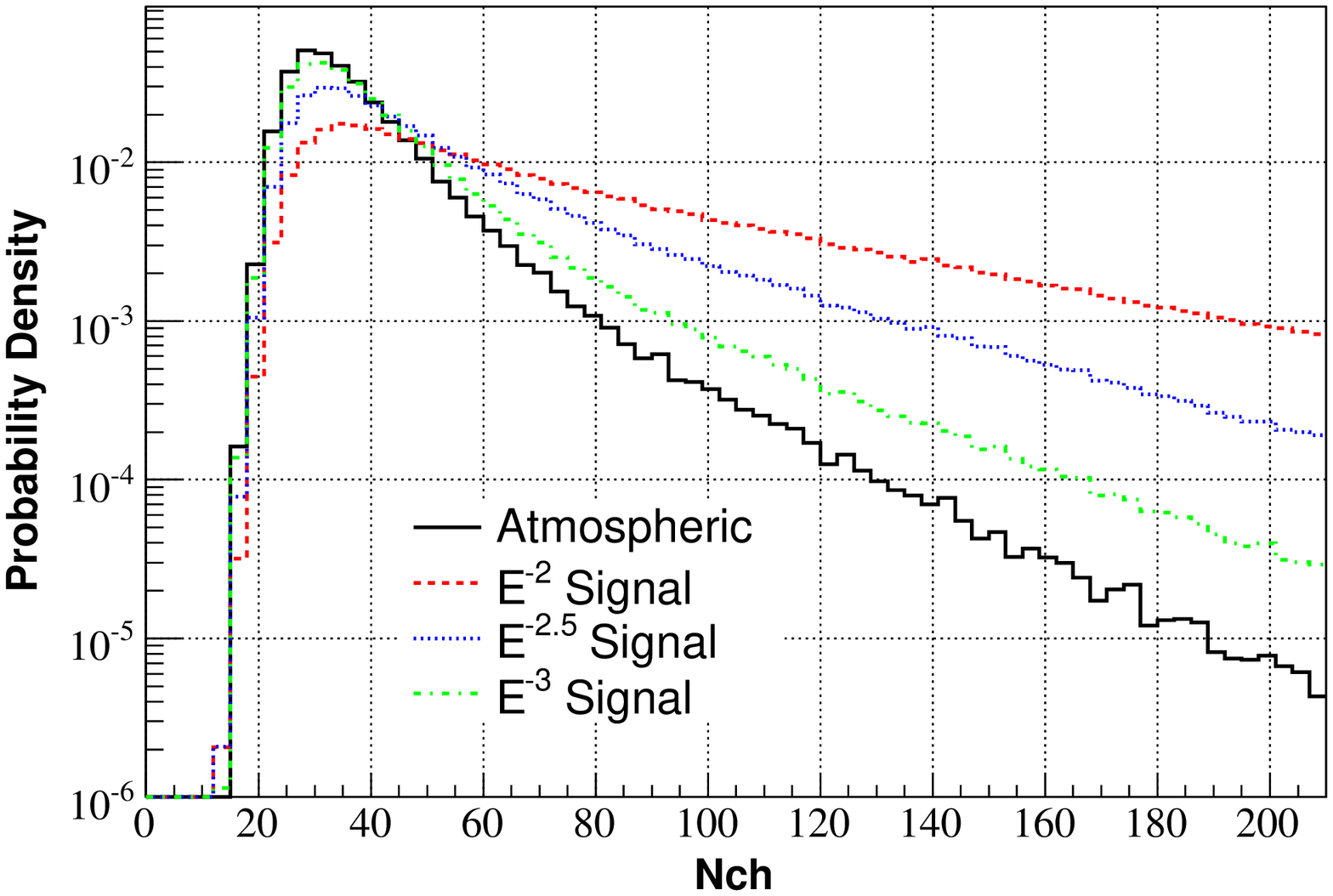}}
\caption{\label{Fig:Edist} Muon energy distributions for four ranges of $N_{ch}$ (top), and simulated $N_{ch}$ distributions for atmospheric
neutrinos \cite{bartol} and E$^{-2}$, E$^{-2.5}$, and E$^{-3}$ power law neutrino spectra (bottom).}
\end{center}\end{figure}

The amount of light deposited in the detector depends strongly on muon energy above $\sim$1 TeV, and thus the number of hit modules ($N_{ch}$)
provides an approximate measure of event energy.  Distributions of muon energy for several ranges of $N_{ch}$ (Fig. \ref{Fig:Edist}) show
the performance of $N_{ch}$ as a muon energy estimator, with a 1$\sigma$ uncertainty in $\log_{10} (E_{\mu}/GeV)$ of 0.65.  Rather than measure event
absolute energy, it is more relevant for a neutrino search to assess the compatibility of an event with expected astrophysical neutrino spectra, assumed
to follow a power law.  From simulations, we tabulate $N_{ch}$ probabilities for spectral indices $1 \le \gamma \le 4$ in bins of 0.01 and for atmospheric
neutrinos \cite{bartol}, shown in Fig. \ref{Fig:Edist}.  This table yields the probability of observing a given $N_{ch}$ value from a source with a power law
energy spectrum relative to observing the value from background atmospheric neutrinos.

\subsection{Maximum Likelihood Method}

\begin{figure}\begin{center}
\mbox{\includegraphics[width=8.6cm]{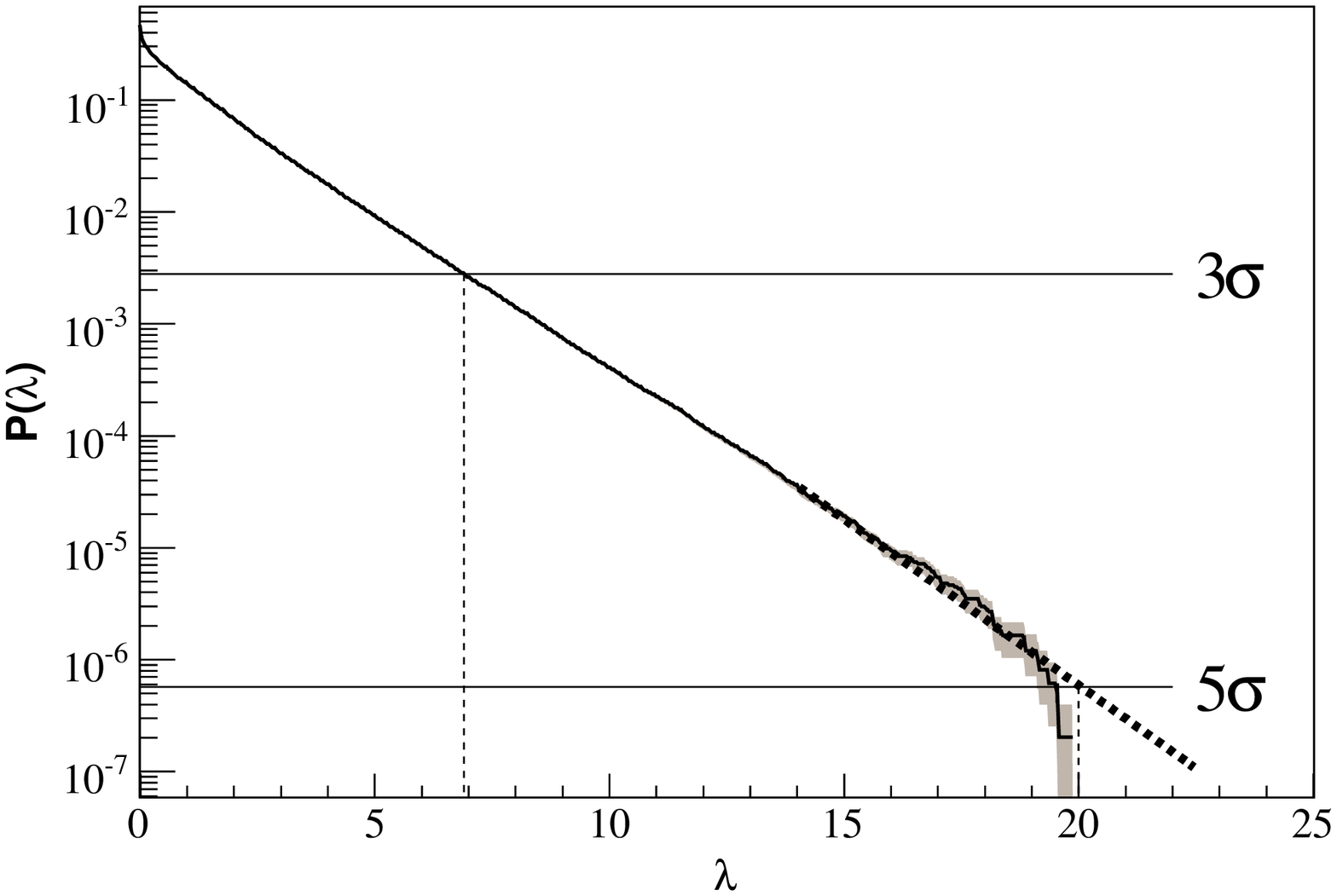}}
\mbox{\includegraphics[width=8.6cm]{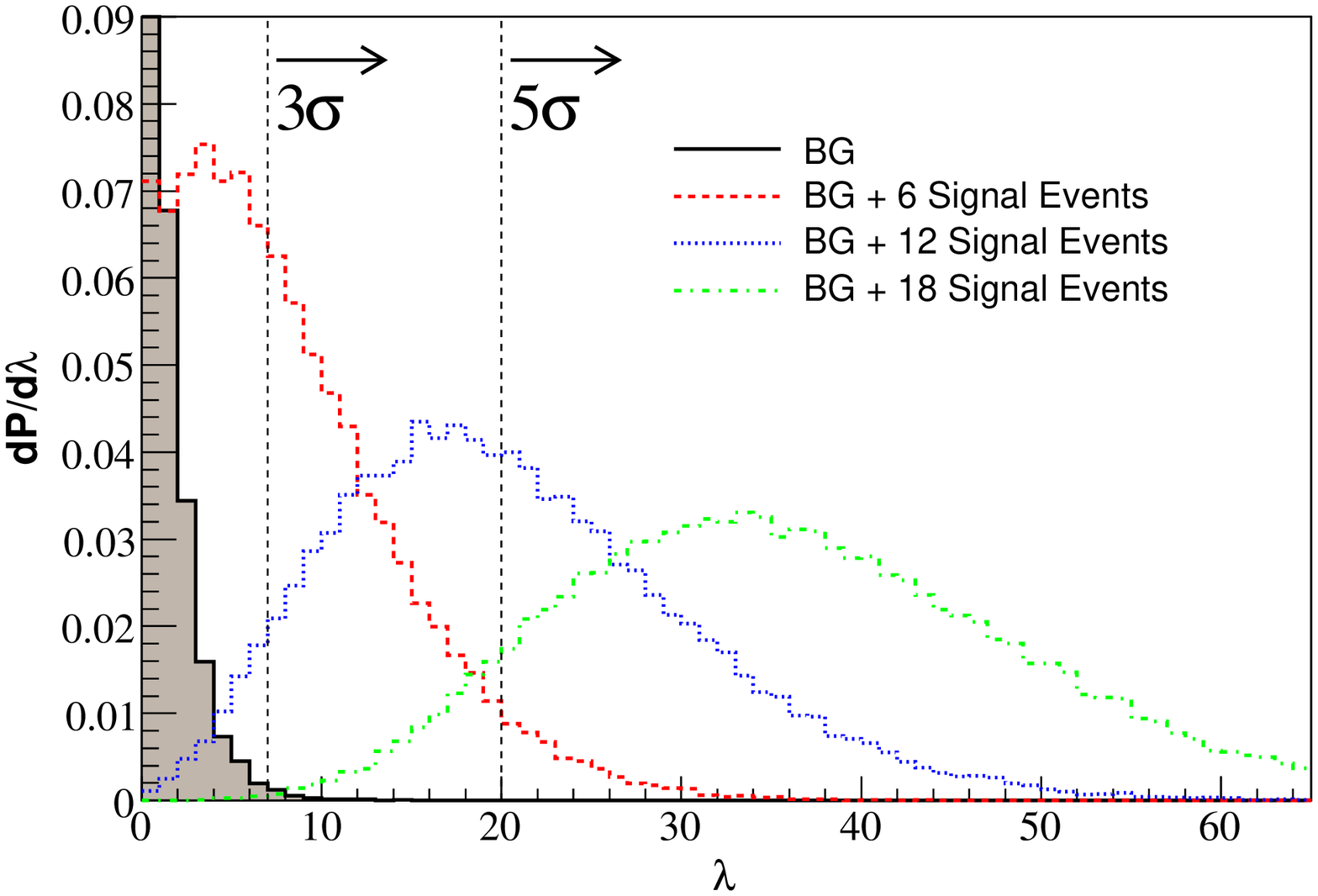}}
\caption{\label{Fig:Lambda1} Integral distribution of the test statistic for background at $\delta$=42.5$^{\circ}$ with 3$\sigma$ and 5$\sigma$
thresholds indicated and statistical uncertainty shaded in gray (top), and distribution of the test statistic for background and 6, 12, and 18
added $E^{-2}$ signal events at $\delta$=42.5$^{\circ}$ (bottom).}
\end{center}\end{figure}

For a source with position $\vec x_s$, giving $n_s$ events against a background of $N-n_s$ events, the probability
density is
\begin{equation}
\frac{n_s}{N}\mathcal{S} + (1 - \frac{n_s}{N})\mathcal{B},
\end{equation}
where $\mathcal{S}$ and $\mathcal{B}$ are the probability densities for signal and background, respectively.
The likelihood function is
\begin{equation}
\mathcal{L} = \prod_{i=1}^{N} \Big(\frac{n_s}{N}\mathcal{S}_{i} + (1 - \frac{n_s}{N})\mathcal{B}_i \Big),
\end{equation}
where $i$ runs over the selected events.  Events are assumed to have an angular error distributed according to a Gaussian given by
the event angular uncertainty $\sigma_i$, and signal events are assumed to follow a power law energy spectrum with spectral index $\gamma$.  The signal
probability density for an event at $\vec x_i$ is
\begin{equation}
\mathcal{S}_{i} = \frac{1}{2\pi \sigma_i^2}e^{-\frac{|\vec x_i - \vec x_s|^2}{2\sigma_i^2}}P(N_{ch,i}|\gamma),
\end{equation}
where $|\vec x_i - \vec x_s|$ is the angular distance between the event and assumed source position.
In practice, we only include events with declinations $\pm 8^{\circ}$ of the source declination since events outside this band have extremely
low signal probabilities, and we set $N$ to be the number of events in this declination band.  The background probability over this band
is roughly constant and given by
\begin{equation}
\mathcal{B}_{i} = \frac{P(N_{ch,i}|\phi_{atm})}{\Omega_{band}}.
\end{equation}
The likelihood $\mathcal{L}$ is maximized (again, $-\log \mathcal{L}$ is numerically minimized) with respect to $n_s$ and $\gamma$, resulting in best
fit signal strength $\hat{n}_s$ and spectral index $\hat{\gamma}$.  The data are then compared to the null, background-only hypothesis ($n_s = 0$)
to determine relative compatibility.  We use as our test statistic
\begin{equation}
\lambda = -2\cdot \log\Big(\frac{\mathcal{L}(n_s = 0)}{\mathcal{L}(\hat{n}_s, \hat{\gamma})}\Big),
\end{equation}
Larger values of $\lambda$ reject the null hypothesis with increasing confidence, shown in Fig. (\ref{Fig:Lambda1}).
The significance of a particular value of $\lambda$ is
determined by comparing the obtained value to the distribution of test statistic values at the same location from data randomized in right ascension,
and we denote as $p$ the fraction of randomized data sets with higher test statistic values.  This method, by using unbinned event-by-event energy and
directional discrimination, improves the sensitivity 
to $E^{-2}$ neutrino fluxes by more than 30\% relative to the previous method \cite{icicrc} using angular bins.

\section{Search for Point Sources in the Northern Sky}

We first apply the search to a predefined list of 26 energetic galactic and extragalactic objects, including many TeV gamma ray sources.
For each source location, we compute the value of the unbinned search test statistic $\lambda$.  Flux upper limits are computed from the
test statistic using Feldman-Cousins unified ordering \cite{fc}.  Systematic uncertainties are incorporated into the limit calculation using
the method of Conrad {\it et al.} \cite{conrad} as modified by Hill \cite{garysys}.  We estimate the total systematic uncertainty in our event
rate expectations for $E^{-2}$ fluxes to be 17\%, summarized in table \ref{Tab:Unc}. Significant contributions include the absolute sensitivity of
optical modules (9\%), neutrino interaction cross section (8\%), bias in event selection between data and simulation (7\%), and photon propagation
in the ice (5\%), determined by detailed detector studies presented in \cite{amandalim}.
\begin{table}[t]
\begin{tabular}{lc}
\hline\hline
Source & Magnitude\\
\hline
Neutrino cross section and rock density & $\pm 8$ \% \\
Optical module sensitivity & ${}^{+2}_{-9}$ \%\\
Photon propagation & $\pm 5$ \% \\
Event selection bias & ${}^{+0}_{-7}$ \%\\
Event reconstruction bias & ${}^{+0}_{-7}$ \%\\
Other known sources & $<4$ \% \\ & \\
\bf{Total} & ${}^{\bf{+10}}_{\bf{-17}}$ \bf{\%}\\
\hline\hline
\end{tabular}
\caption{\label{Tab:Unc}
Systematic errors in event rate expectations for point sources with $E^{-2}$ energy spectra.}
\end{table}
Additionally, we evaluate bias in reconstruction accuracy by comparing distributions of event angular
resolution estimates ($\sigma_i$) with those from point source simulations. We find the angular resolution estimates in simulation are typically 8\% smaller,
and adjusting our simulated point spread by this factor results in flux limits 7\% higher.
Other known sources of systematic uncertainty, including uncertainties in optical module timing resolution and the search method, total less than 4\%.
Limits on $\nu_{\mu}$ + $\nu_{\tau}$ fluxes at 90\% confidence level and
chance probabilities ($p$) are shown in table \ref{Tab:1}.  Limits on $\nu_{\mu}$ fluxes alone correspond to half these values.
The highest significance is found for Geminga with $p$ = 0.0086.  The
probability of obtaining $p$ $\le$ 0.0086 by chance for at least one of 26 sources is 20\% and is therefore not significant.

\begin{table}[t!!!]
\begin{tabular}{lr@{\hspace{0.3cm}}r@{\hspace{0.3cm}}rccc}
\hline\hline
Candidate
&$\delta$($^{\circ}$)
&$\alpha$($h$)
&$\Phi_{90}$
&$p$
&$\Psi$($^{\circ}$)
&N\\
\hline

3C 273 & 2.05 & 12.49 & 8.71 & 0.086 & 2.1 & 3\\
SS 433 & 4.98 & 19.19 & 3.21 & 0.64 & 2.2 & 1\\
GRS 1915+105 & 10.95 & 19.25 & 7.76 & 0.11 & 2.3 & 8\\
M87 & 12.39 & 12.51 & 4.49 & 0.43 & 2.3 & 3\\
PKS 0528+134 & 13.53 & 5.52 & 3.26 & 0.64 & 2.3 & 0\\
3C 454.3 & 16.15 & 22.90 & 2.58 & 0.73 & 2.3 & 5\\
Geminga & 17.77 & 6.57 & 12.77 & 0.0086 & 2.3 & 2\\
Crab Nebula & 22.01 & 5.58 & 9.27 & 0.10 & 2.3 & 7\\
GRO J0422+32 & 32.91 & 4.36 & 2.75 & 0.76 & 2.2 & 3\\
Cyg X-1 & 35.20 & 19.97 & 4.00 & 0.57 & 2.1 & 3\\
MGRO J2019+37 & 36.83 & 20.32 & 9.67 & 0.077 & 2.1 & 7\\
4C 38.41 & 38.14 & 16.59 & 2.20 & 0.85 & 2.1 & 4\\
Mrk 421 & 38.21 & 11.07 & 2.54 & 0.82 & 2.1 & 3\\
Mrk 501 & 39.76 & 16.90 & 7.28 & 0.22 & 2.0 & 6\\
Cyg A & 40.73 & 19.99 & 9.24 & 0.095 & 2.0 & 3\\
Cyg X-3 & 40.96 & 20.54 & 6.59 & 0.29 & 2.0 & 8\\
Cyg OB2 & 41.32 & 20.55 & 6.39 & 0.30 & 2.0 & 8\\
NGC 1275 & 41.51 & 3.33 & 4.50 & 0.47 & 2.0 & 4\\
BL Lac & 42.28 & 22.05 & 5.13 & 0.38 & 2.0 & 2\\
H 1426+428 & 42.68 & 14.48 & 5.68 & 0.36 & 2.0 & 3\\
3C66A & 43.04 & 2.38 & 8.06 & 0.18 & 2.0 & 6\\
XTE J1118+480 & 48.04 & 11.30 & 5.17 & 0.50 & 1.8 & 3\\
1ES 2344+514 & 51.71 & 23.78 & 5.74 & 0.44 & 1.7 & 2\\
Cas A & 58.82 & 23.39 & 3.83 & 0.67 & 1.6 & 2\\
LS I +61 303 & 61.23 & 2.68 & 14.74 & 0.034 & 1.5 & 5\\
1ES 1959+650 & 65.15 & 20.00 & 6.76 & 0.44 & 1.5 & 5\\

\hline\hline
\end{tabular}
\caption{\label{Tab:1}
        Flux upper limits for 26 neutrino source candidates:  Source declination, right ascension,
        90\% confidence level upper limits for $\nu_{\mu} + \nu_{\tau}$ fluxes with $E^{-2}$ spectra ($E^{2}\Phi_{\nu_{\mu} + \nu_{\tau}} \le
        \Phi_{90} \times 10^{-11}\,\mathrm{TeV}\,\mathrm{cm}^{-2}\,\mathrm{s}^{-1}$) over the energy range 1.9 TeV to 2.5 PeV,
        pre-trials significance, median angular resolution of primary neutrino, and number of events inside a cone centered on
        the source location with radius equal to the median point spread.  Since event energy is an important factor in the analysis,
        the number of nearby events does not directly correlate with pre-trials significance.}
\end{table}

\begin{figure}[t]\begin{center}
\mbox{\includegraphics[width=8.6cm]{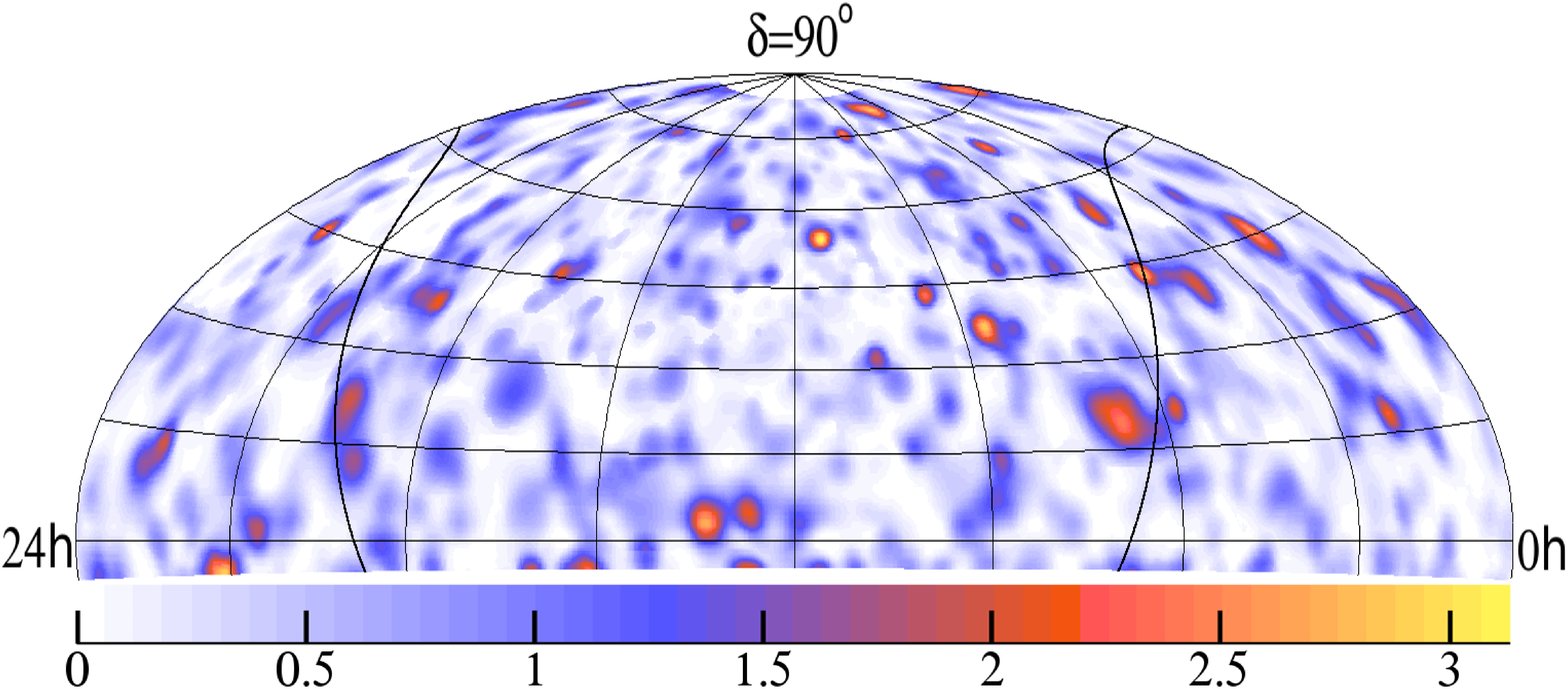}}
\mbox{\includegraphics[width=8.6cm]{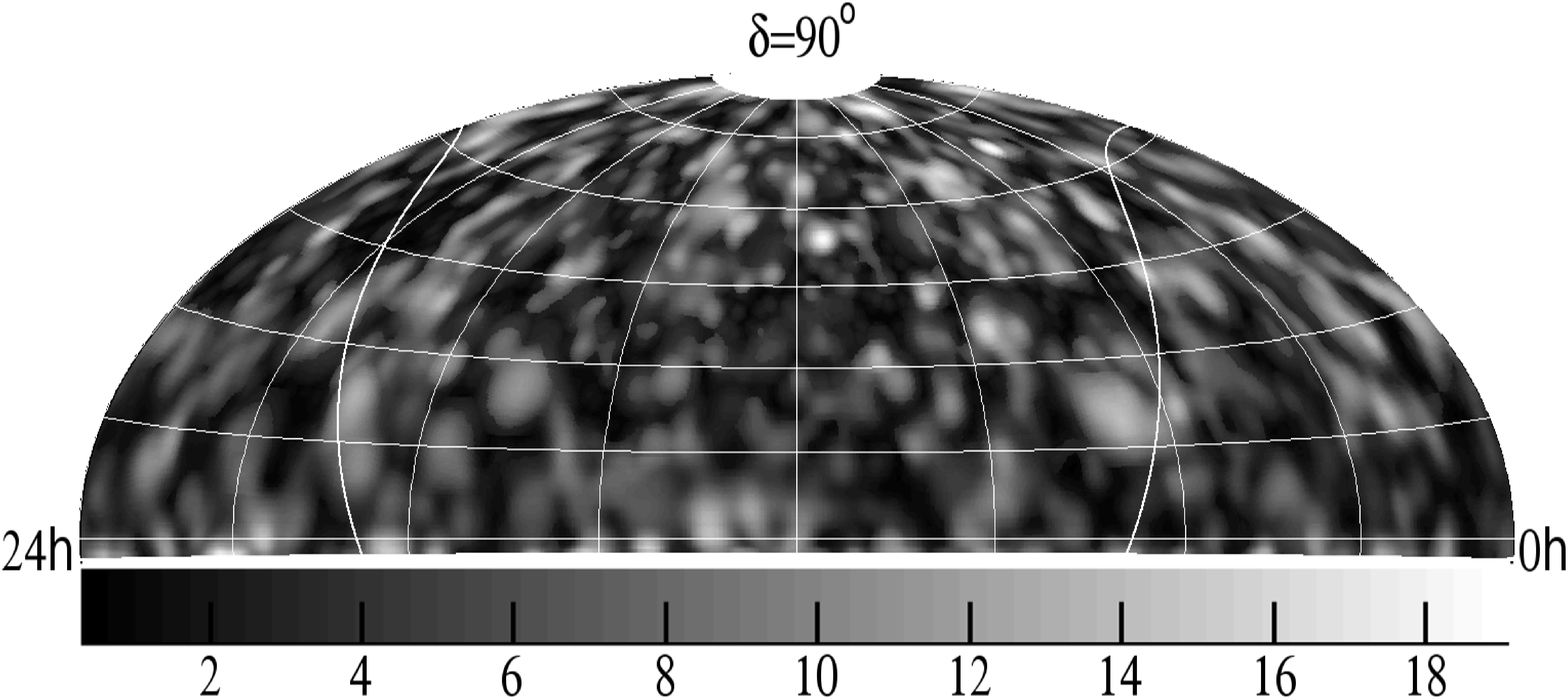}}
\mbox{\includegraphics[width=8.6cm]{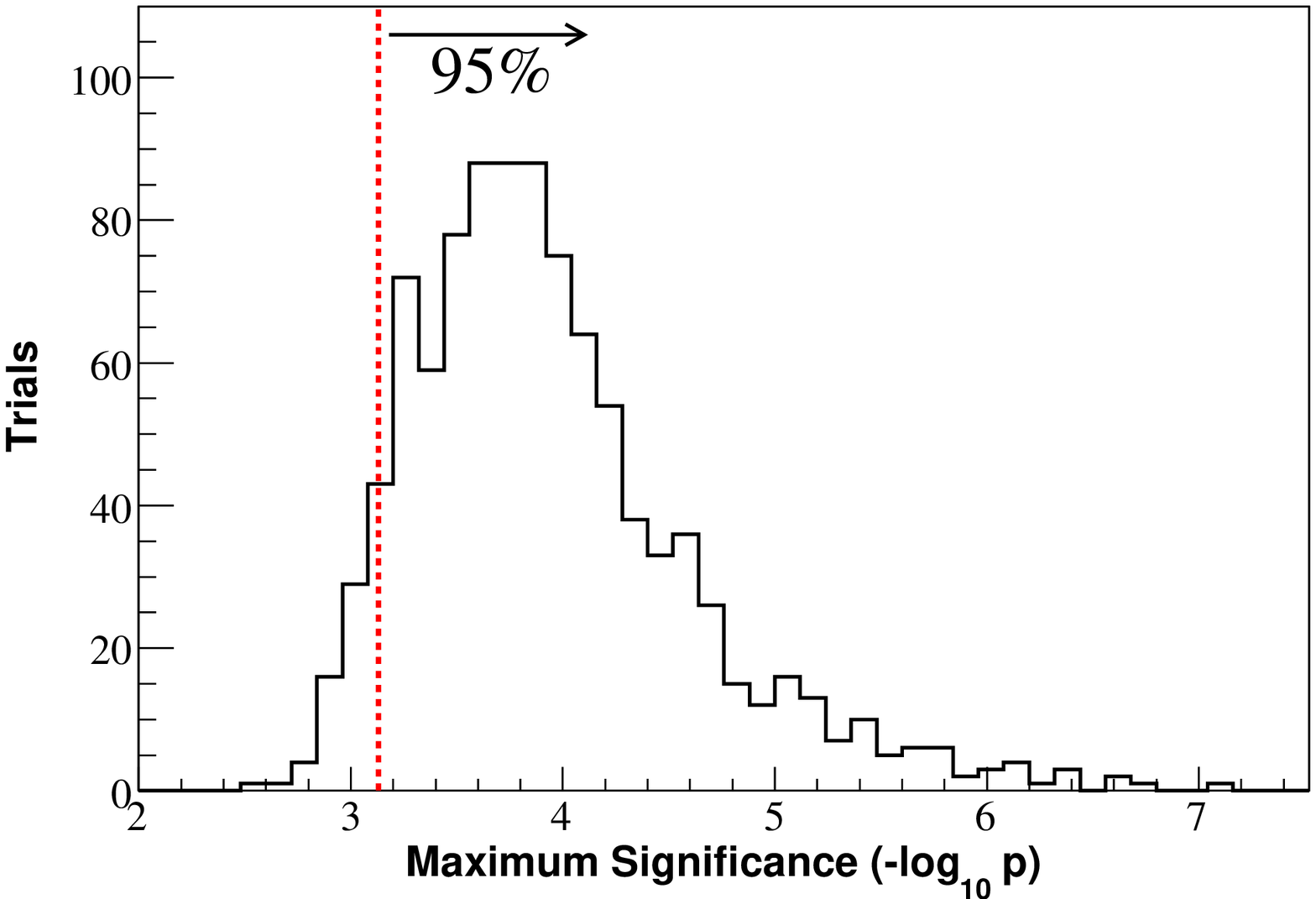}}
\caption{\label{Fig:sig} Sky map of significances ($-\log_{10} p$) obtained in the full-sky search excluding trial factors (top), sky map
of $\nu_{\mu}$ + $\nu_{\tau}$ 90\% confidence level flux upper limits for an $E^{-2}$ energy spectrum (10$^{-11}$ TeV cm$^{-2}$ s$^{-1}$) over
the energy range 1.9 TeV to 2.5 PeV (middle), and the distribution of maximum significances for 1000 randomized sky maps, with
the obtained significance $p$ = 7.4$\times$10$^{-4}$ dotted (bottom).}
\end{center}\end{figure}

\begin{figure}[thb]\begin{center}
\mbox{\includegraphics[width=8.6cm]{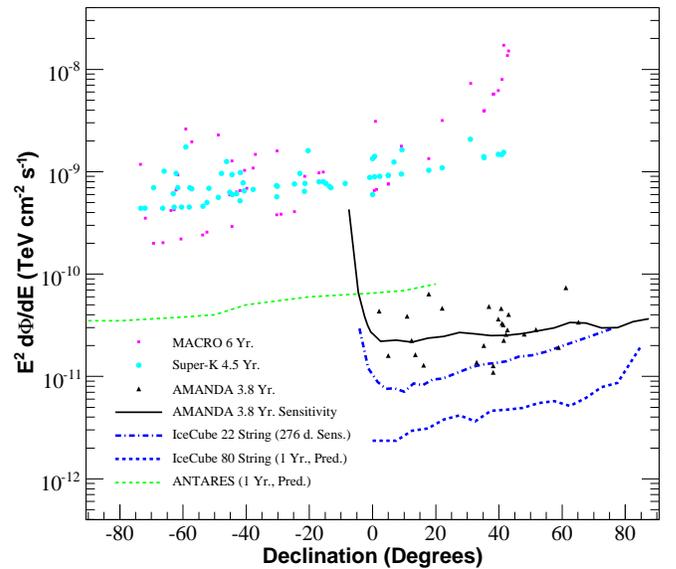}}
\caption{\label{Fig:Limits} Limits on an $E^{-2}$ muon neutrino flux for sources in table \ref{Tab:1} from this work (triangles), limits from
MACRO \cite{macrolim}, and Super-K \cite{superklim},
$E^{-2}$ $\nu_{\mu}$ sensitivity for this work and the IceCube 22-string analysis, and predicted sensitivity for
ANTARES \cite{aguilar} and IceCube.  Our $\nu_{\mu} + \nu_{\tau}$ limits are divided by 2 for comparison with limits on only $\nu_{\mu}$.}
\end{center}\end{figure}

We then apply the search to declinations $-5^{\circ} < \delta < 83^{\circ}$ on a 0.25$^{\circ} \times$ 0.25$^{\circ}$ grid.  The region
above declination $83^{\circ}$ is left to a dedicated search for WIMP annihilation at the center of the Earth \cite{wimpearth}.
For each grid point, we similarly compute a flux limit
and significance (Fig. \ref{Fig:sig}).  We find a maximum pre-trial significance of $p$ = 7.4$\times$10$^{-4}$
at $\delta$ = 54$^{\circ}$, $\alpha$ = 11.4h.
We account for the trial factor associated with the all sky search by comparing the maximum pre-trial significance to the
distribution of maximum pre-trial significances obtained from sky maps randomized in right ascension.
We find 95\% of sky maps randomized in right ascension have a maximum significance of at least $p$ = 7.4$\times$10$^{-4}$ (Fig. \ref{Fig:sig}).
Sensitivity and flux limits are summarized in Fig. \ref{Fig:Limits}.

In the Northern Sky, the galactic TeV gamma ray sources observed by Milagro \cite{milagrogal}
are promising candidates for observation with neutrino telescopes \cite{kappes,beacom}.
We improve our ability to detect a weak signal from this class of objects by $\sim\sqrt{N}$ by
combining $N$ sources of similar strength, with less improvement if one source is much stronger than average.
We include five of eight sources and source candidates observed by Milagro with significance above 5$\sigma$ before considering trial factors,
including four regions near Cygnus and one near the Equator.  We add a hot spot near $\delta$ = 1$^{\circ}$, $\alpha$ = 19h \cite{abdothesis},
which may be associated with a large neutrino flux if confirmed as a source \cite{kappes}.
We exclude the three regions with pulsar-wind nebula counterparts, C3, C4, and the Crab Nebula, which are considered weaker
candidates for significant hadron acceleration \cite{kappes}.  We adapt a method
developed by HiRes \cite{hiresstacking} to perform our maximum likelihood search
simultaneously for all six source locations, resulting in the slightly modified likelihood function
\begin{equation}
\mathcal{L} = \prod_{i=1}^{N} \Bigg(\frac{1}{6}\cdot \frac{n_s}{N}\sum_{j=1}^{6} \mathcal{S}_{i}^j + (1 - \frac{n_s}{N})\mathcal{B}_{i}\Bigg),
\end{equation}
where $\mathcal{S}_{i}^j$ is the signal probability density of the $i^{th}$ event evaluated for
the $j^{th}$ source.  Significance is again computed by comparing the obtained test statistic value to
the distribution obtained from data randomized in right ascension.  We observe a small excess with a chance probability of 20\%.
The 90\% confidence level upper limit obtained on the mean $\nu_{\mu}$ flux per source is 9.7 $\times$ 10$^{-12}$ TeV cm$^{-2}$ s$^{-1}$.

Finally, we search for groups of neutrino sources and extended regions of neutrino emission by scanning for
correlations of events at all angular distances up to 8$^{\circ}$.  We perform the search over a range of energy
thresholds, using the number of modules hit as an energy parameter.  For each threshold in angular distance and number
of modules hit, we count the number of event pairs in the data and compare with the distribution of pairs from data
randomized in right ascension to
compute significance.
%\begin{figure}[t!!]\begin{center}
%\mbox{\includegraphics[width=8.6cm]{FIG_9.eps}}
%\caption{\label{Fig:Auto} Significance ($-\log_{10} p$) of the observed number of event
%pairs with respect to thresholds on angular separation and number of modules hit.}
%\end{center}\end{figure}
The highest obtained significance is $p$ = 0.1
with a threshold of 146 modules hit and 2.8$^{\circ}$ angular separation, where we observe two event pairs.  The probability of
observing this maximum significance by chance is 99\%.  Since four separate analyses are
performed on the data, the probability of obtaining at least one significant result is increased.  The most significant result obtained
has a chance probability of 20\%, and the binomial probability of obtaining this chance fluctuation in at least one of the four
analyses is 59\% and not significant.

\section{Conclusions}

We have analyzed 3.8 years of AMANDA-II data and found no evidence of high energy neutrino point sources.
We place the most stringent limits to date on astrophysical point source fluxes.  IceCube \cite{icecube} is a next-generation 
neutrino telescope at the South Pole scheduled for completion in 2011 with eighty 60-module strings instrumenting $\sim$1 km$^3$ of ice.  Analysis of
data recorded during 2007-2008 with the first 22 strings has improved the AMANDA-II sensitivity by a factor of 2.  Currently 59 strings
are operating, and with continued construction IceCube will achieve an angular resolution of better than one degree and an order of magnitude
improvement over the AMANDA-II sensitivity within a few years.

\section{Acknowledgments}

We acknowledge the support from the following agencies: U.S. National Science Foundation--Office of Polar Programs, U.S. National Science Foundation--Physics Division, University of Wisconsin Alumni Research Foundation, U.S. Department of Energy and National Energy Research
Scientific Computing Center, Louisiana Optical Network Initiative (LONI) grid computing resources, Swedish Research Council, Swedish Polar Research Secretariat, Knut and Alice Wallenberg Foundation (Sweden), German Ministry for Education and Research (BMBF), Deutsche Forschungsgemeinschaft (DFG), (Germany), Fund for Scientific Research (FNRS-FWO), Flanders Institute to encourage scientific and technological research in industry (IWT), Belgian Federal Science Policy Office (Belspo), and the Netherlands Organisation for Scientific Research (NWO); M. Ribordy acknowledges the support of the SNF (Switzerland); A. Kappes and A. Gro{\ss} acknowledge support by the EU Marie Curie OIF Program;
M. Stamatikos is supported by an NPP Fellowship at NASA--GSFC administered by ORAU.

\end{document}